\documentclass[aip,rsi,amsmath,amssymb,reprint,]{revtex4-1}

\usepackage{graphicx}
\usepackage{dcolumn}
\usepackage{bm}

\begin{document}

\title{Compact and lightweight 1.5 $\mu$m lidar with a multi-mode fiber coupling free-running InGaAs/InP single-photon detector}

 \author{Chao Yu}
 \altaffiliation{C. Yu and J. Qiu contributed equally to this work.}
 \affiliation{Hefei National Laboratory for Physical Sciences at Microscale and Department of Modern Physics,
 University of Science and Technology of China, Hefei, Anhui 230026, China}
 \affiliation{CAS Center for Excellence in Quantum Information and Quantum Physics,
 University of Science and Technology of China, Hefei, Anhui 230026, China}

 \author{Jiawei Qiu}%
 \altaffiliation{C. Yu and J. Qiu contributed equally to this work.}
 \affiliation{CAS Key Laboratory of Geospace Environment, University of Science and Technology of China, Hefei, Anhui 230026, China}

 \author{Haiyun Xia}
 \email{hsia@ustc.edu.cn}
 \affiliation{CAS Center for Excellence in Quantum Information and Quantum Physics,
 University of Science and Technology of China, Hefei, Anhui 230026, China}
 \affiliation{CAS Key Laboratory of Geospace Environment, University of Science and Technology of China, Hefei, Anhui 230026, China}

 \author{Xiankang Dou}
 \affiliation{CAS Key Laboratory of Geospace Environment, University of Science and Technology of China, Hefei, Anhui 230026, China}

 \author{Jun Zhang}
 \email{zhangjun@ustc.edu.cn}
 \affiliation{Hefei National Laboratory for Physical Sciences at Microscale and Department of Modern Physics,
 University of Science and Technology of China, Hefei, Anhui 230026, China}
 \affiliation{CAS Center for Excellence in Quantum Information and Quantum Physics,
 University of Science and Technology of China, Hefei, Anhui 230026, China}

 \author{Jian-Wei Pan}
 \affiliation{Hefei National Laboratory for Physical Sciences at Microscale and Department of Modern Physics,
 University of Science and Technology of China, Hefei, Anhui 230026, China}
 \affiliation{CAS Center for Excellence in Quantum Information and Quantum Physics,
 University of Science and Technology of China, Hefei, Anhui 230026, China}

\date{\today}

\begin{abstract}

We present a compact and lightweight 1.5 $\mu$m lidar using a free-running single-photon detector (SPD) based on a multi-mode fiber (MMF) coupling InGaAs/InP negative feedback avalanche diode. The ultimate light detection sensitivity of SPD highly reduces the power requirement of laser, whilst the enhanced collection efficiency due to MMF coupling significantly reduces the volume and weight of telescopes.
We develop a specific algorithm for the corrections of errors caused by the SPD and erbium-doped fiber amplifier to extract accurate backscattering signals.
We also perform a comparison between single-mode fiber (SMF) coupling and MMF coupling in the lidar receiver, and the results show that the collection efficiency with MMF coupling is five times higher than SMF coupling.
In order to validate the functionality, we use the lidar system for the application of cloud detection. The lidar system exhibits the ability to detect both the cloud base height and the thickness of multi-layer clouds to an altitude of 12 km with a temporal resolution of 1 s and a spatial resolution of 15 m.
Due to the advantages of compactness and lightweight, our lidar system can be installed on unmanned aerial vehicles for wide applications in practice.

\end{abstract}

\maketitle

\section{Introduction}

Aerosols are the extended colloidal dispersion of solid or liquid particles in a gaseous medium such as smoke, fog, sea salt, soil dust, and combustion products. The occurrence, residence time, and physical and chemical properties of aerosols vary fast in space and time.
For instance, clouds are essentially aerosols consisting of minute liquid droplets and frozen crystals suspended in the atmosphere, which fundamentally influence weather, air travel safety, sunlight illumination and climate.
Lidar (light detection and ranging), including coherent lidar and direct detection lidar, provides an effective technology to detect aerosols with high resolution. Compared with coherent lidar, direct detection lidar can directly detect the intensity of backscattering signals with simple system design and data processing, which is well suited for cloud and aerosol detections.
Commercial direct detection lidars widely use lasers at wavelengths of 532 nm, 910 nm and 1064 nm~\cite{GCC10,LXX15}, due to the fact that mature silicon avalanche photodiodes can be used at these wavelengths.

Since 1.5 $\mu$m laser has advantages of eye-safe, low atmospheric attenuation and low solar background energy, such wavelength is also an ideal candidate for lidar applications. The performance in terms of detection range of 1.5 $\mu$m direct detection lidar is considerably limited by the detection of ultraweak backscattering signals. For cloud detection, the reported ranges using 1.5 $\mu$m lidars are only $\sim$ 6 km~\cite{JJM98,AAR10} and no commercial 1.5 $\mu$m ceilometers have been reported so far. The widely used commercial ceilometer, Vaisala CL31, is operated at 910 nm, which has the maximum detection altitude of 7.6 km, the minimum temporal resolution of 2 s and a weight of 30 kg~\cite{CL31}.

Near infrared free-running single-photon detectors (SPDs) have the ultimate light detection sensitivity, therefore, 1.5 $\mu$m lidars using SPDs can significantly increase the detection range and resolution of aerosols.
There are currently three major free-running SPDs, i.e., superconducting nanowire single-photon detectors (SNSPDs), up-conversion single-photon detectors (UCSPDs), and InGaAs/InP SPDs using negative feedback avalanche diodes (NFADs), as lised in Table~\ref{table1}.
SNSPDs exhibit the best performance~\cite{FVJ13,XHW14,JLS17}, e.g., 93\% photon detection efficiency (PDE), 1 cps dark count rate (DCR), zero afterpulse probability, and 25 Mcps maximum count rates (MCR). Although SNSPDs are widely used for lidar academic researches~\cite{MHC17,JHX17}, the requirement of cryogenic conditions and high cost limit their use for practical applications.
UCSPDs require room temperature condition and also have moderate performance~\cite{HGM15,MHC16,HMC16}, e.g., 20\% PDE, 60 cps DCR and 0.94\% afterpulse probability, however, the internal waveguide requires only single-mode fiber (SMF) coupling for the incident light.

In practical applications, compactness and lightweight are the primary challenges for lidar system, and free-running InGaAs/InP SPD is the most appropriate solution, due to the advantages of small size, multi-mode fiber (MMF) coupling, polarization-independent, and low cost~\cite{XMK11,ZDA12,TCO12,BNT14,JMH15,YS17}.
On one hand, InGaAs/InP SPD is much more compact that SNSPD and UCSPD.
On the other hand, MMF coupling highly enhances the collection efficiency
of backscattering signals distorted by atmospheric turbulence compared with SMF coupling~\cite{YF05}, which, therefore, significantly reduces the volume and weight of telescopes.

In this paper, we present a compact and lightweight 1.5 $\mu$m lidar system using a MMF coupling free-running InGaAs/InP SPD.
We develop an algorithm for the corrections of SPD noises and amplified spontaneous emission (ASE) noises to obtain accurate backscattering signals.
Then we apply the lidar system for cloud detection, and the experimental results show that the lidar system can detect both the cloud base height and the thickness of multi-layer clouds to an altitude of 12 km with a temporal resolution of 1 s and a spatial resolution of 15 m.

\begin{table}[tbp]
\centering
\label{table1}
\caption{Comparison of different free-running SPDs.}
\begin{tabular}{cccc}
\hline
                                & InGaAs/InP SPD        & UCSPD~\cite{HGM15}    & SNSPD~\cite{FVJ13} \\
\hline
Fiber mode                      & SMF/MMF              & SMF                 & SMF/MMF   \\
Polarization            & independent    & dependent      &dependent \\
Temperature             & 223 K                 & 300 K                     &  120 mK       \\
Volume             & compact                 & moderate                     &  bulky       \\
PDE            & 10\%                        & 20\%                        & 93\%               \\
DCR                & 2000 cps                       & 60 cps                      & 1 cps        \\
Afterpulse          & 10\%                      & 0.94\%                      & 0\%         \\
MCR             & 1.6 Mcps                 & 37 Mcps                     &  25 Mcps       \\

\hline
\end{tabular}
\end{table}

\section{System}

\begin{figure}[bp]
\centering
\includegraphics[width=8 cm]{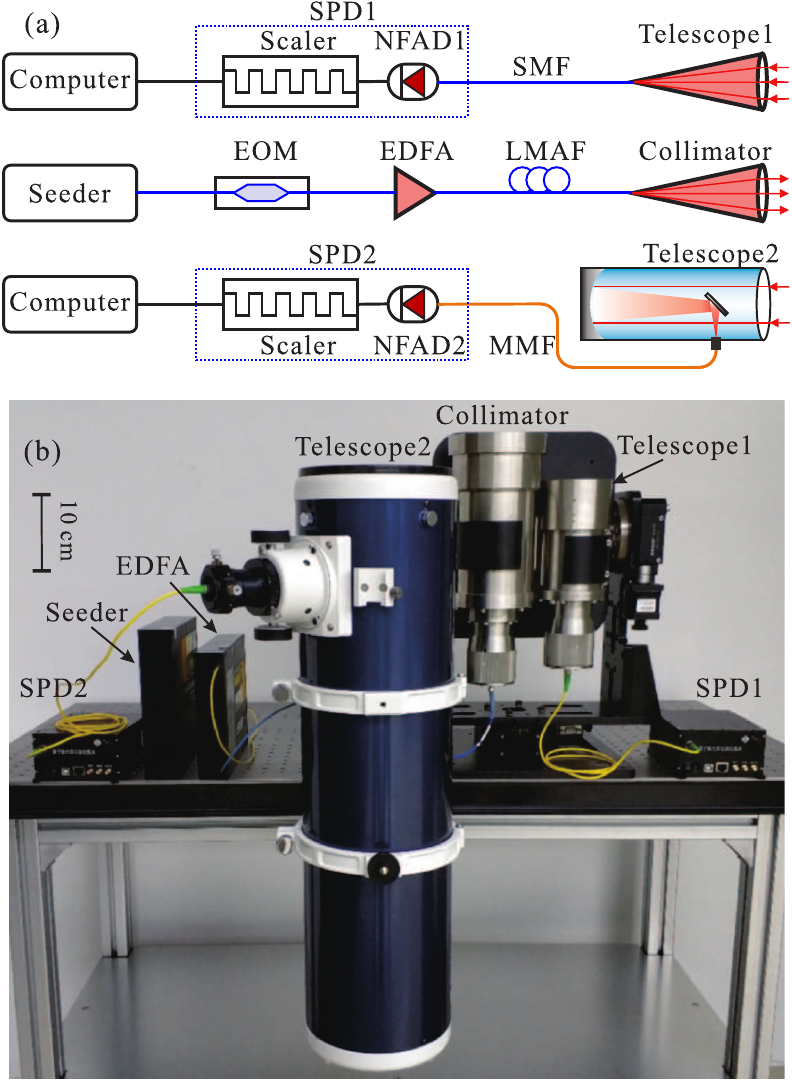}
\caption{Experimental setup (a) and photo (b) of 1.5 $\mu$m lidar using free-running InGaAs/InP SPD.
Telescope 1 is a refracting telescope consisting of 3 quartz spherical lens.
EOM: electro-optic modulator; EDFA: erbium-doped fiber amplifier; LMAF: large-mode-area fiber; SMF: single-mode fiber; MMF: multi-mode fiber; NFAD: negative feedback avalanche diode.}
\label{fig1}
\end{figure}

The setup diagram and photo of 1.5 $\mu$m lidar system are shown in Fig.~\ref{fig1}(a) and Fig.~\ref{fig1}(b), respectively. In order to perform direct comparison, a MMF coupling receiver and a SMF coupling receiver are used simultaneously with the same transmitter in the experiment. In the transmitter, a continuous wave (CW) laser at a wavelength of 1548.1 nm is reshaped to a pulsed laser with 100 ns pulse width and 10 kHz repetition frequency, via an electro-optic modulator (EOM, Photline, MXER-LN-10) with an extinction ratio as high as 35 dB. The weak laser pulses are then amplified to $\sim$30 $\mu$J per pulse by an erbium-doped fiber amplifier (EDFA, Keyopsys, PEFA-EOLA). After passing a large-mode-area fiber (LMAF) and a collimator with 100 mm diameter, the laser pulses are transmitted into atmosphere vertically with a 40 $\mu$rad divergence angle.

The backscattering signals are coupled both to a SMF coupling NFAD with 9 $\mu$m diameter and 0.12 numerical aperture (NA) and to a MMF coupling NFAD (Princeton Lightwave) with 62.5 $\mu$m diameter and 0.275 NA. In order to guarantee that the receiver's field-of-view (FOV) is larger than the transmitter's FOV, the telescope diameters for SMF coupling and MMF coupling are designed to 70 mm with 50 $\mu$rad divergence angle, and 150 mm with 80 $\mu$rad divergence angle, respectively. The backscattering signals are detected by two free-running InGaAs/InP SPDs (SPD1 and SPD2), and the timing information of detection events is transmitted to a computer for real-time processing and corrections. The key parameters of the lidar system are listed in Table~\ref{table2}.

\begin{table}[t]
\centering
\label{table2}
\caption{Parameters of the lidar system.}
\begin{tabular}{lc}
\hline
Parameter                               & Value      \\  
\hline
 Transmitter \\
 \quad Wavelength                      & 1.5 $\mu$m            \\    

 \quad Pulse width                     & 100 ns            \\  

 \quad Pulse energy                    & 30 $\mu$J           \\    

 \quad Repetition rate                 & 10 kHz            \\   

 \quad Beam divergence        & 40 $\mu$rad            \\  

  Receiver1 \\
 \quad Detector        & SMF NFAD \\

 \quad Telescope diameter   & 70 mm \\

 \quad Fiber core diameter  & 9 $\mu$m \\

 \quad Fiber NA             & 0.12 \\

 \quad Beam divergence      & 50 $\mu$rad   \\

 Receiver2 \\
 \quad Detector        & MMF NFAD          \\

 \quad Telescope diameter   & 150 mm \\

 \quad Fiber core diameter  & 62.5 $\mu$m \\

 \quad Fiber NA             & 0.275 \\

 \quad Beam divergence      & 80 $\mu$rad   \\

 System  \\
 \quad Measurement range               & 12 km          \\  

 \quad Spatial resolution                & 15 m            \\

 \quad Temporal resolution            & 1 s            \\

 \quad Weight               & 15 kg \\

\hline
\end{tabular}
\end{table}

Inside the free-running InGaAs/InP SPD, the NFAD is cooled down to 223 K using thermoelectric coolers (TECs, Thermonamic), and the quenching circuit is similar to our previous work~\cite{YS17}. By optimizing the quenching electronics design and the layout of printed circuit board, the dimensions of SPD are reduced down to 100 mm $\times$ 150 mm $\times$ 60 mm. The weight of SPD is $\sim$ 1 kg.
We then characterize the InGaAs/InP SPD using the standard calibration approach~\cite{JMH15}, and optimize the parameters to achieve high count rate that is required for lidar applications. Under the conditions of 223 K and 600 ns hold-off time, the InGaAs/InP SPD exhibits performance of 10\% PDE, 2 kcps DCR, 10\% afterpulse probability and 1.6 Mcps MCR. Further, an internal scaler with 10 ns bin width is implemented to tag the timing information of detection events by a field-programmable gate array (FPGA) in the quenching circuit of SPD.

\section{Results}

\begin{figure}[tbp]
\centering
\includegraphics[width=8 cm]{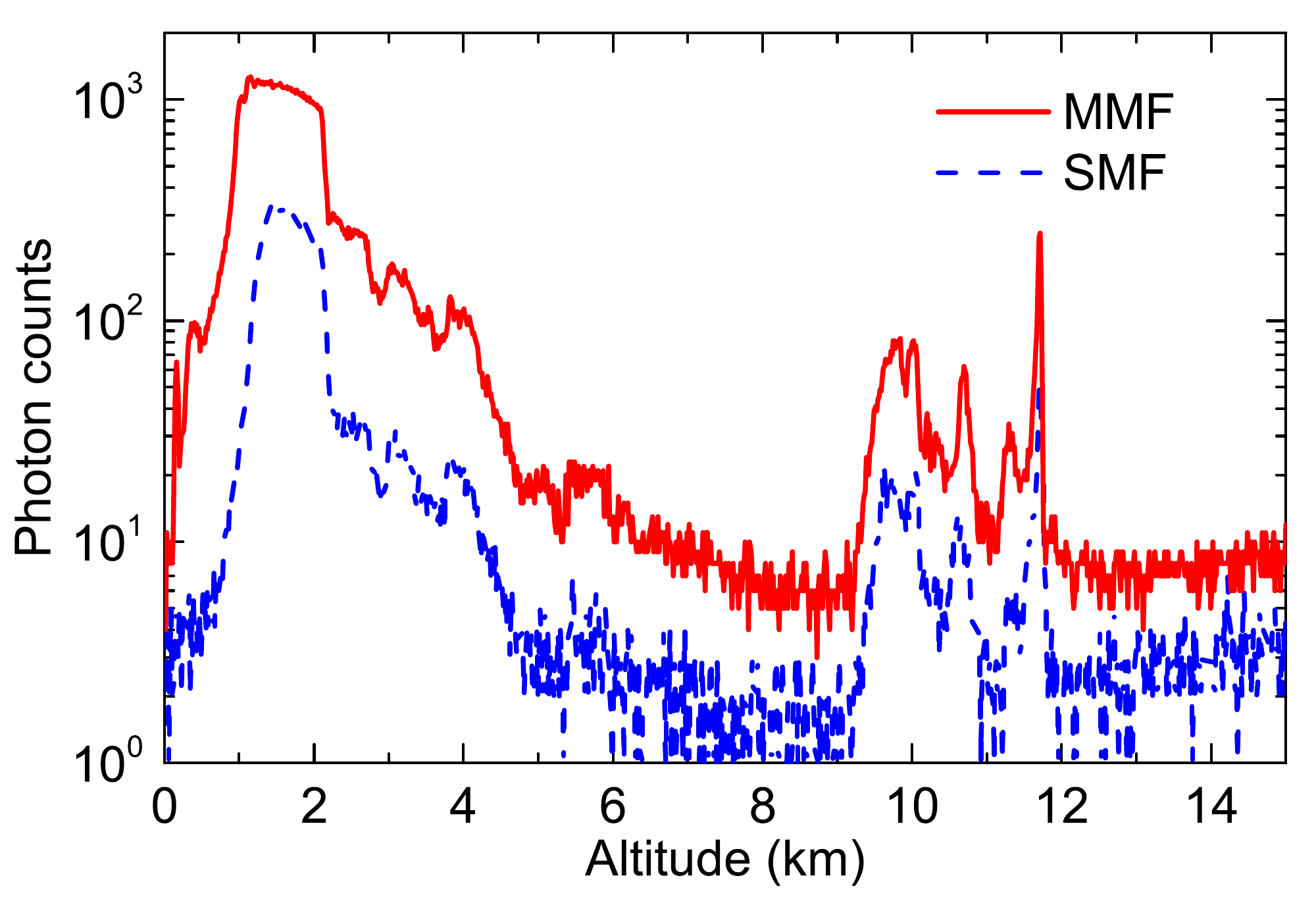}
\caption{Original photon counts as a function of altitude for a vertical cloud detection observation using SMF and MMF receivers with a measurement time of 1 s. The first and the second peaks are due to the backscattering signals of aerosols in the near field and different layers of clouds in the far field, respectively.}
\label{fig2}
\end{figure}

Fig.~\ref{fig2} shows the original data of photon counts versus altitude with the two receivers for a vertical observation of cloud detection.
In the experiment, the parallel-axis telescopes are used. The overlap geometry factor of a biaxial lidar is adopted to suppress the strong backscattering in low altitude, avoiding the saturation of SPD. The geometry factors of the two telescopes are zero at the ground and reach full overlap beyond 1.5 km.
The first peak in the near field is contributed by the backscattering signal of aerosols, and the second peak ranging from 9 km to 12 km is the backscattering signal from different layers of clouds. At very near range, the small difference between two cases is caused by the deviation of overlap factors and different saturation degrees. From Fig.~\ref{fig2}, one can conclude that the backscattering signal intensity using MMF coupling is around five times higher than SMF coupling due to the large diameter and NA of MMF.

To extract backscattering signals with high signal-to-noise ratio (SNR), device imperfections have to be corrected, including the contributions by hold-off time, DCR, afterpulse probability of SPD and the ASE noise of EDFA. The effect of hold-off time is often roughly corrected by
\begin{equation}
\label{R_CR}
R_{CR\_C}(i)=\frac{R(i)}{1-R(i)\tau}-DCR(i),
\end{equation}
where $R(i)$, $DCR(i)$ are the photon count rate and dark count rate at bin $i$, respectively, and $\tau$ is the hold-off time.
However, in cloud lidar applications, due to the existence of sharp backscattering signal edges as shown in Fig.~\ref{fig2},
the hold-off time correction should be as accurate as possible. Generally, from the fact that the detection event occurred at bin $i$ means non-detection events occurred in $\tau$ before bin $i$, one can deduce the following hold-off time correction
\begin{equation}
\label{R_HT}
R_{HT\_C}(i)=\frac{R(i)}{1-\sum_{k=i-\tau}^iR(k)/f}-DCR(i),
\end{equation}
where $f$ is system repetition frequency, i.e., 10 kHz. Eq.\ref{R_HT} is particularly appropriate for fast-varying signal correction, and for slow-varying signal Eq.\ref{R_HT} is equivalent to Eq.\ref{R_CR}.

\begin{figure}[tbp]
\centering
\includegraphics[width=8 cm]{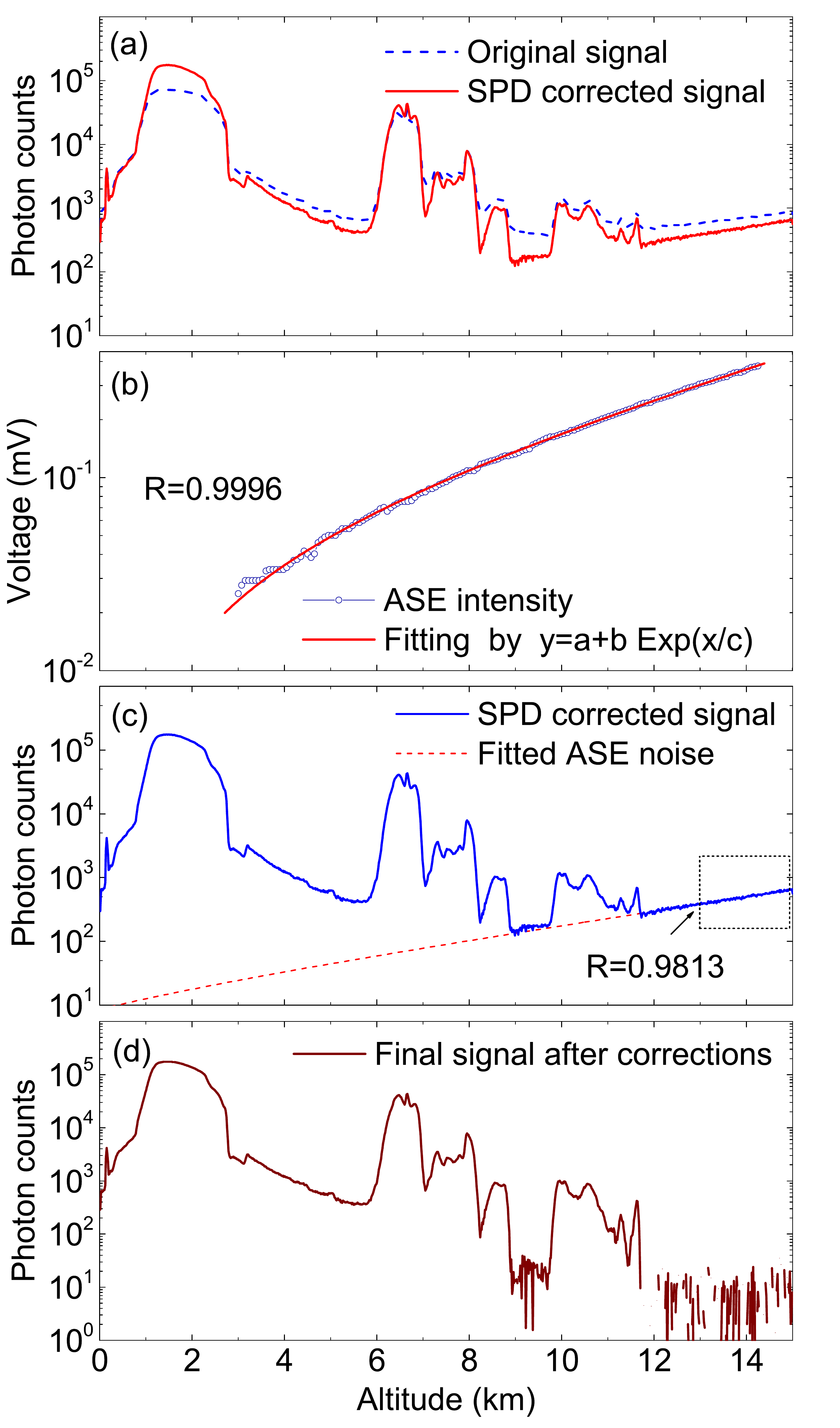}
\caption{Error corrections and cloud backscattering signal extraction. (a) Original signal and the corrected signal after SPD parameter corrections. (b) Measured ASE noise intensity distribution from EDFA as a function of time (corresponding to altitude) by shielding the laser pulses. (c) ASE noise fitting in the backscattering signal. (d) Final extracted backscattering signal.}
\label{fig3}
\end{figure}

Further, the contribution of the afterpulsing effect can corrected as~\cite{YS17}
\begin{equation}
\label{R_AP_C}
R_{AP\_C}(i)=R_{HT\_C}(i)-R_{ap}(i),
\end{equation}
where $R_{ap}(i)$ represents the afterpulse count rate at bin $i$. The detection events in other bins may contribute afterpulse counts at bin $i$, therefore, $R_{ap}(i)$ is calculated by~\cite{YS17}
\begin{equation}
\label{R_AP}
R_{ap}(i)=\sum_jR(j)P(i,j),
\end{equation}
where $P(i,j)$ is the probability of a detection event at bin $j$ inducing an afterpulse count at bin $i$. Such conditional probability consists of three parts, i.e., 1) there is no detection event between bin $j$ and bin $i$; 2) there is no afterpulse between bin $j$ and bin $i$; 3) an afterpulse is created at bin $i$. Then, $P(i,j)$ can be calculated as~\cite{YS17}
\begin{equation}
\label{P(i,j)}
P(i,j)=Exp(-\sum_{k=j}^iR(k)/f)Exp(-\sum_{k=0}^{i-j-1}P_{ap}(k))P_{ap}(i-j),
\end{equation}
where $P_{ap}(i)$ is the afterpulse probability at bin $i$ that can be obtained from the characterized afterpulse probability distribution of SPD. After the above processes, the SPD errors can be corrected, as shown in Fig.~\ref{fig3}(a).

The ASE noise of EDFA is another important error source in the lidar system. After sending a laser pulse, the spontaneous radiation in EDFA increases over time. In the data processing of lidar systems, optical flying time is related to radial distance, e.g., 1 $\mu$s time between the backscattering signal and the outgoing laser pulse corresponds to $\sim$ 150 m distance. Therefore, the intensity of ASE noise increases along the altitude. Fig.~\ref{fig3}(b) plots the measured ASE noise intensity distribution by shielding the laser pulses with an EOM.
The measured intensity at each bin is fitted by
\begin{equation}
\label{V_ASE}
V_{ASE}(i)=a+b Exp(i/c).
\end{equation}

In the lidar system, the transmission telescope is collimated for the wavelength of laser pulses. Due to the wide spectrum characteristic, the ASE noises only in the near range can be collected inside the receiver's FOV.
The ASE count rate is proportional to the intensity of ASE noise. In the experiment, considering the aerosol backscattering signals are negligible within the range from 13 km to 15 km, we use the data in this range to fit the count rate distribution of ASE noise, $R_{ASE}(i)$, according to Eq.~\ref{V_ASE}. The fitted result is shown in Fig.~\ref{fig3}(c). Then, the final extracted signal is given by
\begin{equation}
\label{RC}
R_C(i)=R_{AP\_C}(i)-R_{ASE}(i).
\end{equation}
After performing all the corrections, the SNR of backscattering signal is significantly enhanced, particularly at the range of high altitude, as shown in Fig.~\ref{fig3}(d).

\begin{figure}[tbp]
\centering
\includegraphics[width=8.5 cm]{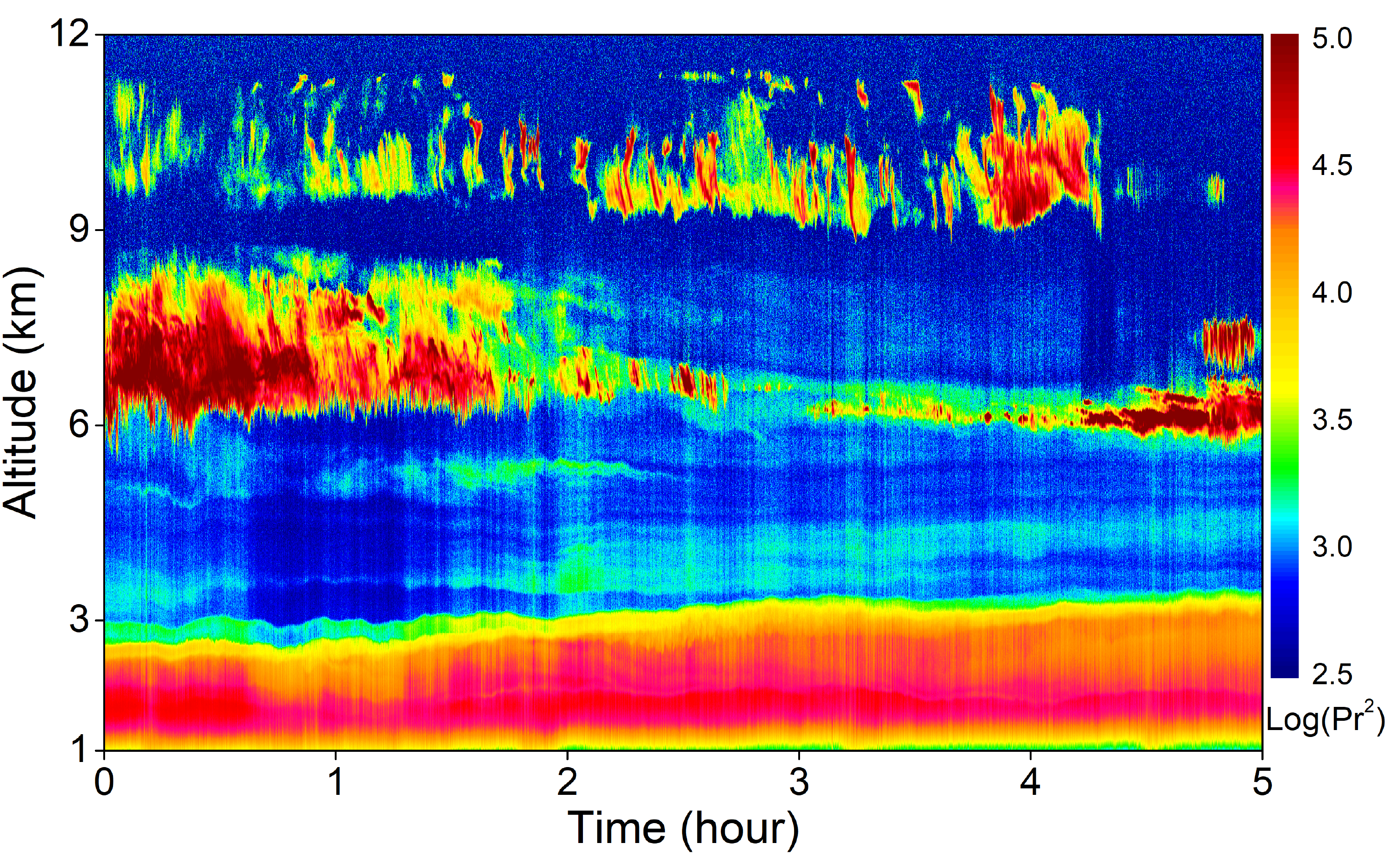}
\caption{Range corrected signal ($\log Pr^2$) of multi-layer clouds using the MMF receiver, observed from 00:00 to 05:00 on March 24, 2018.}
\label{fig4}
\end{figure}

Finally, we perform a continuous observation with the MMF receiver from 00:00 to 05:00 on March 24, 2018 to verify the stability of the lidar system for cloud detection, as shown in Fig.~\ref{fig4}. The altitude of boundary layer is $\sim$3 km during the observation. Below the boundary layer, the backscattering signals are attributed by aerosols. Between 6 km and 10 km, two layers of clouds are clearly detected. During the first hour observation, the lower and upper clouds are simultaneously detected with thickness of $\sim$3 km, $\sim$1.5 km, respectively, which indicates the high sensitivity of the lidar system for multi-layer cloud detection.

\section{Conclusion}

In summary, we have reported a compact and lightweight 1.5 $\mu$m lidar system based on a MMF coupling free-running InGaAs/InP SPD. The MMF coupling can
significantly enhance the collection efficiency of backscattering signal. We have developed a specific algorithm to correct the noises contributed by SPD and EDFA. The lidar system has achieved the detection of multi-layer clouds to an altitude of 12 km with a temporal resolution of 1 s and a spatial resolution of 15 m. Due to compactness and light weight, the lidar system can be deployed on unmanned aerial vehicles for practical uses.

\section*{acknowledgements}

The authors thank M. Shangguan and W.-H. Jiang for useful discussions. This work has been supported by the National Key R\&D Program of China under Grant No.~2017YFA0304004, the National Natural Science Foundation of China under Grant No.~11674307, the Chinese Academy of Sciences, and the Anhui Initiative in Quantum Information Technologies.

\end{document}